\documentclass[twocolumn,showpacs,prl]{revtex4}
\usepackage{amssymb}
\usepackage{graphicx}
\usepackage{dcolumn}
\usepackage{bm}
\usepackage{amsmath}

\begin{document}

\title{Suppression of inhomogeneous broadening in rf spectroscopy of optically
trapped atoms}
\author{Ariel Kaplan, Mikkel Fredslund Andersen, and Nir Davidson}
\affiliation{Department of Physics of Complex Systems,\\
Weizmann Institute of Science, Rehovot 76100, Israel}

\begin{abstract}
We present a novel method for reducing the inhomogeneous frequency
broadening in the hyperfine splitting of the ground state of
optically trapped atoms. This reduction is achieved by the
addition of a weak light field, spatially mode-matched with the
trapping field and whose frequency is tuned in-between the two
hyperfine levels. We experimentally demonstrate the new scheme
with $^{85}$Rb atoms, and report a 50-fold narrowing of the rf
spectrum.
\end{abstract}

\maketitle

The time available for measuring an atomic transition in a
perturbation-free environment has been substantially increased by
the achievement of ultracold atomic samples. Using cold atoms in
atomic fountains, measurements of the 9.2 GHz ''clock'' transition
in Cesium were performed with measurement times as long as 450
ms\cite{Kasevich89}. A possible way to increase the measurement
time beyond the practical limit imposed by the height of a
fountain is to use optically trapped atoms\cite{Chu86,Grimm00}.
Far-off-resonance optical traps (FORTs)\cite{Miller93}, which are
based on the conservative dipole force created by cycles of
absorption and stimulated emission, are possible candidates for
these measurements. The great disadvantage of a trap, is that the
trapping potential acts also as a perturbation for the atomic
level, and in particular an optical trap introduces a relative ac
Stark shift of the hyperfine levels\cite{Davidson95}, which
results in a systematic shift in the clock frequency measurement.
When ensemble-averaged, the spatial dependence of the potential
also results in a inhomogeneous broadening of the transition, and
hence in a loss of atomic coherence at a much faster rate than the
spontaneous photon scattering rate\cite{Davidson95}. These effects
can be reduced by increasing the trap detuning
\cite{Davidson95,Adams95,Miller93,Takekoshi96} and by using
blue-detuned optical traps, in which the atoms are confined mainly
in the dark\cite{Davidson95,Friedman02}. However, the residual
frequency shifts are still the main limiting factor for precision
spectroscopy in optical traps.

ac Stark shifts from the trapping beams are also detrimental to
achieving high phase space densities in optical traps, since they
shift the atom's resonance frequency away from the cooling beams
frequency \cite {Gordon80,Kuppens00}. Recently, it was shown that
in some cases a FORT laser frequency can be chosen to couple the
states involved in cooling to some other states, in order to
suppress the frequency shift of the cooling transition and allow
simultaneous trapping and Doppler cooling\cite{Katori99}. In this
way, a phase-space density exceeding 0.1 was achieved.

In this letter, we demonstrate a method for reducing the
inhomogeneous broadening in the spectroscopic measurement of the
hyperfine splitting of the ground state of optically trapped
atoms. This reduction is achieved by the addition of a very weak
light field (the so-called compensating beam), spatially
mode-matched with the trapping field and whose frequency is tuned
between the two hyperfine levels. An experimental realization of
the new scheme is presented with $^{85}$Rb atoms, and a reduction
by a factor 50 is reported. This method can be applied to red- or
blue-detuned FORTs, and hence can be used as an additional way to
further increase the long atomic coherence times of the latter.

A ground state $\left| g_{i}\right\rangle $ of an atom exposed to a light
field with intensity $I$ and frequency $\omega$, is ac Stark shifted by an
amount given by:
\begin{equation}
\Delta E_{i}=\frac{3\pi
c^{2}\gamma}{2\omega_{0}^{3}}I\mathbf{\times}\sum
\frac{c_{ij}^{2}}{\delta_{ij}}  \label{eq_lightshift}
\end{equation}
where $\gamma$ is the natural width of the
transition\cite{Grimm00}. The summation takes into account the
contributions of the different coupled excited levels $\left|
e_{j}\right\rangle $, each with its respective
transition coefficient $c_{ij}$, and detuning $\delta_{ij}=\omega-%
\omega_{ij} $.

\begin{figure}[tp]
\begin{center}
\includegraphics[width=3in]{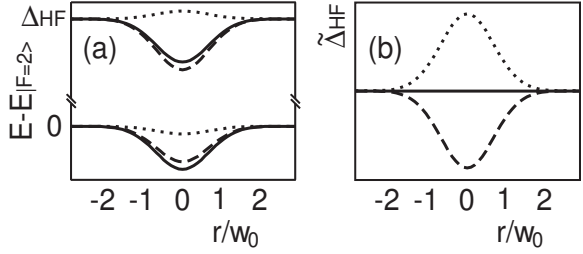}
\end{center}
\caption{Ground level energies (a) and energy difference (b) of
atoms trapped in a focused gaussian beam. When exposed to the
trapping laser, the two hyperfine levels have a different ac Stark
shift (dashed line). An additional weak laser, detuned to the
middle of the hyperfine splitting, creates an ac Stark shift
(dotted line) such that the total amount of light shift (full
line) is identical for both hyperfine levels.}
\label{fig_complevels}
\end{figure}

Specifically for the D2 line in $^{85}$Rb, the fine structure of
the excited state ($\Delta_{F}^{\prime}\sim7$ THz), and the
hyperfine structure of the ground state ($ \Delta _{HF}\sim3$ GHz)
and excited state ($\Delta_{HF}^{\prime}$, tens of MHz)
obey $\Delta_{F}^{\prime}\gg%
\Delta_{HF}\gg\Delta _{HF}^{\prime}$. For linearly polarized light
in the vicinity of this line, and as long as the detuning of the
light is large as compared to the excited state hyperfine
splitting $\Delta_{HF}^{\prime}$, a general result can be derived
from Eq. \ref{eq_lightshift} for the shift of a ground state with
total angular momentum $F$, exposed to a light field $I\left(
\mathbf{r}\right)$:
\begin{equation}
\Delta E\left( \mathbf{r}\right) =\frac{\pi c^{2}\gamma}{\omega_{0}^{3}}%
\frac{I\left( \mathbf{r}\right) }{\delta_{F}} \label{deltaE}
\end{equation}
where $\delta_{F}=\omega-\omega_{F}$ is the detuning of the laser
from the $ \left| 5S_{1/2},F\right\rangle \rightarrow\left|
5P_{3/2}\right\rangle $ transition. Note, that Eq. \ref{deltaE} is
a reasonable approximation even for a detuning $\delta_{F}$
comparable with the optical frequency $\omega_{0}$ \cite{Adams95}.
We are interested in rf spectroscopy, where the energy difference
between the two ground state hyperfine levels, $\left|
F=2\right\rangle $ and $\left| F=3\right\rangle $, is measured. In
the presence of the light this energy difference is modified by:
\begin{equation}
\widetilde{\Delta}_{HF}\left( \mathbf{r}\right) -\Delta_{HF}=
\frac{\pi c^{2}\gamma}{\omega_{0}^{3}}\frac{\Delta_{HF}}{\delta^{2}}\left[ \frac{1}{1-\left( \frac{%
\Delta_{HF}}{2\delta}\right) ^{2}}\right] I\left(
\mathbf{r}\right) \label{eq_newhf}
\end{equation}
where $\widetilde{\Delta}_{HF}\left( \mathbf{r}\right) $ is the
spatially dependent hyperfine splitting in the presence of the
light and $\delta \triangleq(\delta_{F=2}+\delta_{F=3})/2$
measures the laser detuning from the center of the ground state
hyperfine splitting. Equation \ref{eq_newhf} indicates that for
$\left| \delta\right| >\frac{ \Delta_{HF}}{2}$, i.e. for $
\delta_{F=2} $ and $\delta_{F=3}$ both positive (or both
negative), the ground state energy splitting is always reduced by
the presence of a light field (See Fig. \ref{fig_complevels}).
When the detuning is "between" the hyperfine levels, $\left|
\delta\right| <\frac{\Delta_{HF}}{2}$, the energy splitting is
enlarged.

For a FORT with detuning $ \delta \gg\frac{\Delta_{HF}}{2}$ Eqs.
\ref{deltaE},\ref{eq_newhf} yield:
\begin{equation}
\widetilde{\Delta}_{HF}\left( \mathbf{r}\right) -\Delta_{HF}\approx\left( \frac{\Delta_{HF}}{%
\delta}\right) U\left( \mathbf{r}\right)
\end{equation}
where $U\left( \mathbf{r}\right) $ is the spatially dependent
dipole potential that forms the trap, in frequency units. The fact
that the relative ac Stark shift is $\Delta_{HF}\diagup\delta$
smaller than the dipole potential, is the main motivation for
using FORTs for precision spectroscopy. For example in Ref.
\cite{Davidson95} the relative ac Stark shifts were only
$\Delta_{HF}\diagup\delta\approx 2\cdot10^{-4}$ times the dipole
potential. Note that, however small, this relative ac Stark shift
is still much larger than the spontaneous photon scattering rate
\cite{Davidson95}.

For a trapped atomic ensemble, $I\left( \mathbf{r}\right) $ will
cause a shift in the ensemble averaged ground state hyperfine
energy splitting,
$\left\langle\widetilde{\Delta}_{HF}-\Delta_{HF}\right\rangle$. In
addition, the spatial dependence of $ I\left( \mathbf{r}\right) $
will result in an inhomogeneous broadening,
$\sqrt{\left\langle\widetilde{\Delta}_{HF}^{2}\right\rangle-
\left\langle\widetilde{\Delta}_{HF}\right\rangle^{2}} $, which
depends also on the atoms' temperature\cite{Nodynamics}. For
example, for a thermal ensemble of atoms with $\frac{3}{2}k_{b}T$
kinetic energy in an harmonic trap, we have $\left\langle
\widetilde{\Delta}_{HF}-\Delta_{HF}\right\rangle =\left(
\frac{\Delta_{HF}}{\delta}\right) \left[
U_{0}+\frac{3}{2}k_{b}T\right] $ and
$\sqrt{\left\langle\widetilde{\Delta}_{HF}^{2}\right\rangle-
\left\langle\widetilde{\Delta}_{HF}\right\rangle^{2}} =\left(
\frac{\Delta_{HF}}{\delta}\right) \sqrt{\frac{3}{2}}k_{b}T$ ,
where $U_{0}$ is the dipole potential at the trap's bottom
\cite{Kaplan02}.

In order to cancel these shifts, we introduce an additional laser
beam, with intensity $I'\left( \mathbf{r}\right)$ and frequency
between the resonant frequencies of the two ground state hyperfine
levels, say in the middle, i.e. $-\Delta_{HF}/2$ and
$+\Delta_{HF}/2$ from the lower and higher hyperfine level
respectively\cite{Nomiddle}. For a FORT, the total shift is
obtained by adding the shifts from the trap and the compensating
beam,
\begin{equation}
\widetilde{\Delta}_{HF}\left( \mathbf{r}\right) -\Delta_{HF}=\frac{\pi c^{2}\gamma\Delta_{HF}}{%
\omega_{0}^{3}}\left[ \frac{I\left( \mathbf{r}\right)
}{\delta^{2}-\left( \frac {\Delta_{HF}}{2}\right)
^{2}}-\frac{I'\left( \mathbf{r}\right) }{\left(
\frac{\Delta_{HF}}{2}\right) ^{2}} \right] \label{Newhfs}
\end{equation}

If the compensating beam is spatially mode-matched with the trap beam, i.e. $%
I'\left( \mathbf{r}\right) =\eta\times I\left( \mathbf{r}\right)
$, then a complete cancellation of the inhomogeneous broadening
will occur for
\begin{equation}
\eta= \frac{(\frac{\Delta_{HF}}{2})^{2}}{\delta^{2}-\left(
\frac{\Delta_{HF}}{2}\right)
^{2}}\approx(\frac{\Delta_{HF}}{2\delta})^{2} \label{Etha}
\end{equation}

Fig. \ref{fig_complevels} shows the shift of the hyperfine levels
(a) and  the hyperfine energy difference (b) caused by the
trapping beam (dashed line) and compensating beam (dotted line).
In the presence of both beams, the levels are shifted by the same
amount (full line). As a specific example, with a FORT detuned by
5 nm we have $\eta\approx3.6\times10^{-7}$. Hence, with a typical
FORT power of 50 mW, the required compensating beam power is 20
nW\cite{noted1}.

Note, that the dipole potential created by the compensating beam
is $U'\left( \mathbf{r}\right)
=\pm \frac{1}{2}\frac{\Delta _{HF}}{\delta}U\left( \mathbf{r}%
\right) $ for atoms in the upper an lower hyperfine level,
respectively, and hence is negligible when compared with the
potential of the FORT. Moreover, the photon scattering rate from
the compensating beam $\gamma'_{s}$ is given by $\hbar
\gamma'_{s}\approx2\gamma U'/\Delta _{HF}$, which can be also
written as $\hbar \gamma'_{s}\approx\frac{\gamma }{\delta}U\left(
\mathbf{r}\right)\approx\hbar\gamma_{s}$. Hence, the scattering
rate from the nearly resonant compensating beam, is of similar
magnitude to that of the far-of-resonance trapping beam,
$\gamma_{s}$.

We implement the proposed scheme with a FORT (a red-detuned
gaussian beam \cite{Chu86}), created by focusing a 50 mW laser,
detuned 5 nm below resonance, to a waist of $W_{0}=50\mu m$,
resulting in a potential depth of $U_{0}\approx 200E_{rec}$
($E_{rec}$ is the recoil energy) and oscillation frequencies of
$\nu _{r}= 2.3\times 10^{3}$ Hz and $\nu _{z}= 13$ Hz in the
radial and axial directions, respectively. An additional laser,
with frequency locked close to the middle of the ground state
hyperfine splitting, is combined with the FORT laser. To achieve
optimal spatial mode-match, both lasers are coupled into a
polarization-conserving single-mode optical fiber, and the fiber's
output is passed through a polarizer and focused into the vacuum
chamber. Two servo loops are used to control and stabilize the
power of the lasers: The first one ensures a $1\%$ stability of
the trap laser. More importantly, for complete compensation of the
relative ac Sark shifts, a second servo loop ensures a $0.1\%$
stability of the power ratio $\eta$ throughout the entire duration
of the experiment\cite{Gratings}.

The loading procedure is similar to that described in
\cite{Ozeri99}. Briefly, the FORT is loaded by overlapping it with
a compressed $^{85}$Rb magneto-optical trap (MOT). The MOT beams
are shut off after 650 ms of loading, 50 ms of compression, and 5
ms of polarization gradient cooling, leaving $\sim10^{5}$ confined
atoms with a temperature of $\sim10 \mu K$.

\begin{figure}[tp]
\begin{center}
\includegraphics[width=2.8in]{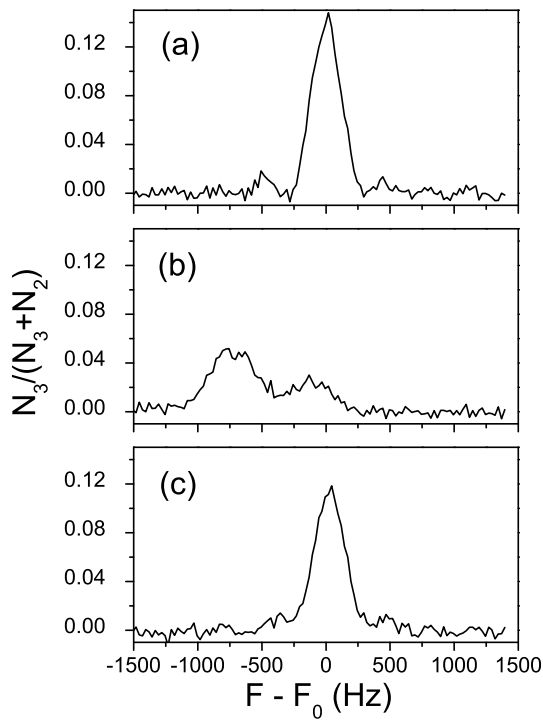}
\end{center}
\caption{Rabi spectrum of the hyperfine splitting of $^{85}$Rb,
with a 3 ms pulse. (a) Spectrum of free falling atoms.
$F_{0}\approx3035732439$ Hz is the free-atoms line center. (b)
Rabi spectrum of trapped atoms, showing a shift in the line center
and a broadening. (c) Spectrum of trapped atoms, with an
additional compensating beam. The addition of the weak
compensating beam, nearly cancels the shift and broadening of the
spectrum. Note that since the population of the four $\left| F=2,
m\neq0\right\rangle$ states is included in $N_{3}/(N_{2}+N_{3})$,
a value of 0.2 represents the maximal possible signal (a $\pi$
pulse) for the $\left| F=2, m=0\right\rangle \rightarrow
\left|F=3, m=0\right\rangle$ transition.} \label{fig_spectrum3}
\end{figure}

We perform a Rabi spectroscopy measurement on the trapped atoms by
driving the ground state $\left| F=2,m_{F}=0\right\rangle
\rightarrow $ $\left| F=3,m_{F}=0\right\rangle $ transition, which
is insensitive to magnetic fields, to the first order. A bias
magnetic field of $\sim $ 80mG is applied parallel to the FORT's
polarization axis and to the rf magnetic field direction, in order
to Zeeman shift the magnetic sensitive $m_{F}\neq 0$ levels out of
resonance with the rf pulse. A typical sequence is as follows:
First, the atoms are prepared in the $ \left| F=2\right\rangle $
ground state by turning on the MOT beams, without a repump
beam\cite{Ozeri99}, for 1 ms. Then, an rf pulse in applied at a
variable frequency, using an Anritsu 69317B Signal Generator
locked to a high stability oscillator. The intensity and duration
of the pulse are adjusted to maximize the $\left| F=3\right\rangle
$ population when on-resonance ($\pi $ pulse condition). Following
the rf pulse, $N_{3}$ (the population in the $\left|
F=3\right\rangle $ level) is measured by detecting the
fluorescence during a short pulse of a laser beam resonant with
the cycling transition $\left| 5S_{1/2},F=3\right\rangle
\rightarrow \left| 5P_{3/2},F=4\right\rangle $. The population of
the $\left| F=2\right\rangle $ level ($N_{2}$) is then measured by
turning on the repumping beam (which is resonant with $\left|
5S_{1/2},F=2\right\rangle \rightarrow \left|
5P_{3/2},F=3\right\rangle $) and applying an additional detection
pulse. The normalized signal $N_{3}/(N_{2}+N_{3})$ is insensitive
to shot-to-shot fluctuations in atom number as well as slow
drifting fluctuations of the detection laser frequency and
intensity\cite{Khaykovich99}.

\begin{figure}[tp]
\begin{center}
\includegraphics[width=3.1in]{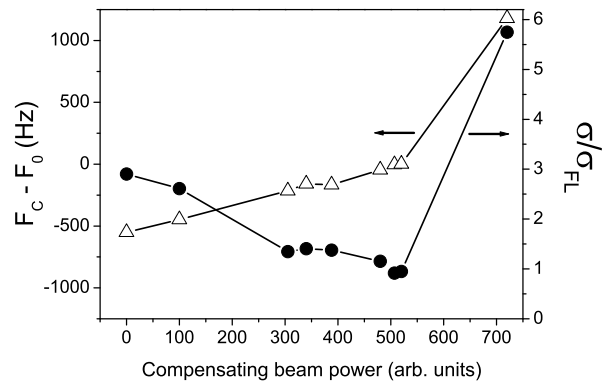}
\end{center}
\caption{Trapped atoms line center ($F_{c}$, $\vartriangle$) and
rms width ($\sigma$, $\bullet$) as a function of compensating beam
power, for a 3 ms pulse. ($\sigma_{FL}\approx110$ Hz is the
Fourier limited $\sigma$). The spectrum width is minimized to a
Fourier limited value, at a compensating beam intensity which
corresponds also to a minimal shift from the free-atoms line
center $F_{0}$.} \label{fig_fwhm}
\end{figure}

Figure \ref{fig_spectrum3} shows results for the Rabi spectrum
with a 3 ms long $\pi$ pulse. A constant background resulting from
spontaneous Raman scattering \cite{Cline94} is substracted. The
spectrum of free-falling atoms shows no inhomogeneous broadening
and a rms width, $\sigma$, which is Fourier limited to $110Hz$. A
shift in the peak frequency ($-756Hz$), and a broadening of the
line (to $\sigma=320Hz$) are seen in the spectrum of trapped
atoms, in fair agreement with the calculated trap depth and atomic
temperature. This inhomogeneous broadening is not significantly
affected by the duration of the pulse \cite{Nodynamics}. The
addition of the weak compensating beam, nearly cancels the
broadening of the spectrum, as well as its shift from the
free-atom line center.

Figure \ref{fig_fwhm} shows the measured rms width and shift of
the trapped atoms as a function of compensating beam power, again
for a 3 ms rf pulse. The spectrum width is minimized to the
Fourier broadening limit at a compensating beam power which
corresponds also to a minimal shift from the free-atoms line
center.

\begin{figure}[tp]
\begin{center}
\includegraphics[width=2.8in]{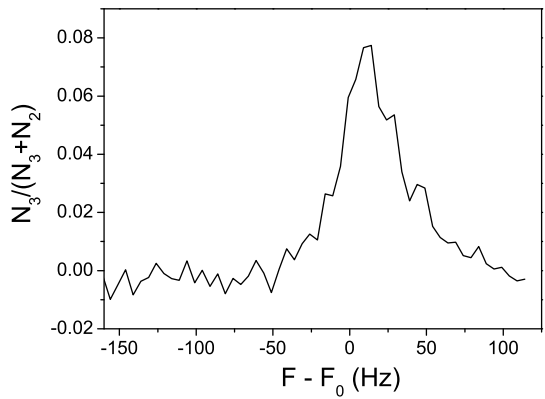}
\end{center} \caption{Rabi spectrum of the hyperfine splitting of
$^{85}$Rb, with a 25 ms pulse. A Fourier limited
$\sigma\approx13Hz$ is measured.} \label{fig_spectrum25}
\end{figure}

Figure \ref{fig_spectrum25} shows the measured rf spectrum for a
25 ms long pulse. A measurement of free atoms with this pulse
length is not possible in our setup since the atoms fall due to
gravity, and leave the interaction region. A Fourier limited
$\sigma=13Hz$ is measured, representing a 25-fold reduction in the
line broadening, as compared to the line broadening of trapped
atoms. We performed a similar measurement with a 50 ms pulse, and
observed a nearly Fourier limited width (a 50-fold narrowing), at
the expense of a much larger spontaneous photon scattering and
hence a smaller signal\cite{Background5}. For even larger
measurement times spontaneous photon scattering prevents further
narrowing of the line.

We measure the spin relaxation rate\cite{Cline94,Ozeri99} to be
$\sim3\times10^{-3} s^{-1}$  for atoms trapped in the FORT. The
addition of the compensating beam induces an increase of only
$\sim20\%$.

In summary, we perform an rf spectroscopy measurement of the
hyperfine splitting of the ground state of optically trapped
atoms. We demonstrate a novel scheme for eliminating the
trap-induced inhomogeneous broadening of the transition, by adding
a weak ''compensating'' laser, spatially mode-matched with the
trapping laser and with a proper detuning and intensity. Despite
being tuned close to resonance, this laser induces a negligible
change in the dipole potential, and does not considerably increase
the spontaneous scattering rate. With the suppression of
inhomogeneous broadening, the atomic coherence time is now limited
by the much smaller spontaneous scattering time.

Whereas in a conventional optical trap the ac Stark shift of the
line center strongly depends on the temperature of the atoms,
which may drift considerably, in the compensated trap the
suppression of the line shift is equally effective for all
temperatures. Hence, it provides a mean of achieving a higher
stability of the line center than that achieved by simply
stabilizing the trapping laser detuning and intensity. For
relative spectroscopic measurements, such as the proposed
measurement of the electron's permanent electric dipole moment
(EDM)\cite{Chin01}, only stability (and not absolute accuracy) of
the line center is of importance. For example, for a 10 $\mu$K
deep YAG-laser trap, and a compensating beam with a 15 KHz (time
averaged) frequency stability, locking the relative intensity
between both beams to a $1:10^{-5}$ stability, will result in
$\sim10^{-14}$ stability of the rf line center.

Finally, a weak compensating beam, spatially mode-matched with the
trapping beam and properly tuned near resonance between the upper
level of a laser cooling transition and another excited level, can
suppress spatially dependent frequency shifts of the cooling
transition, and allow simultaneous trapping and cooling with more
flexibility than the single frequency method of Ref.
\cite{Katori99}.

This work was supported in part by the Israel Science Foundation,
the Minerva Foundation, and the United States-Israel Binational
Science Foundation. MFA acknowledges help from the Nachemsohn
Dansk-Israelsk Studienfond.

\end{document}